\documentclass{PoS}
\usepackage{natbib}
\usepackage{graphicx}
\usepackage{epsfig}

\PoS{PoS(BDMH2004)058}

\title{Bending instabilities at the origin of persistent warps: a new
  constraint on dark matter halos}

\ShortTitle{Bending instabilities at the origin of persistent warps}

\author{\speaker{Y. Revaz} and D. Pfenniger\\
        Geneva Observatory, Switzerland\\
        E-mail: \email{yves.revaz@obs.unige.ch},
	        \email{daniel.pfenniger@obs.unige.ch}}

\abstract{
  Based on N-body simulations, we show that realistic galactic disks
  are subject to bending instabilities of fire-hose type when the
  disks are substantially self-gravitating, that is, if they contain
  dark matter distributed in the disk.  Depending on the degree of
  instability, S and U-shaped, as well as asymmetric warps are
  generated. In some cases, the warp may last several galactic
  rotations, particularly when the instability is marginal.  Since the
  bending instability is very sensitive to the disk flattening, the
  fractions of dark matter distributed in the disk and in the dark
  halo are constrained.  For a Milky Way like galaxy the extended dark
  halo can not exceed 30-40\% of the total mass within 35\,kpc if the
  Milky Way warp results from a bending instability.  This mode of
  warping provides a unified picture of spiral galaxies, where bars,
  spiral arms and warps result all from disk gravitational
  instabilities, radial or transverse, which are constantly
  regenerated by the dissipative gas component.}

\FullConference{Baryons in Dark Matter Halos\\
                5-9 October 2004\\
                Novigrad, Croatia}

\begin{document}

\section{Introduction}

Since the first observation of the Milky Way warp at the end of the
fifties, warped galaxies have represented a challenge for
astrophysicists \citep[see for example][for a review]{binney92}.
Several hypotheses like normal modes, accretion events, or dynamical
friction have been proposed to explain their ubiquity.  Most of the
explanations assume that galactic disks are embedded in a dominant
spheroidal dark halo.  However, there are now several observations
(e.g. of NGC 2915) which can be best explained with a substantial
amount of dark matter in the outer galactic disks (Masset \& Bureau
2003), in agreement with cold and clumpy gas dark matter models
distributed like HI disks \citep{pfenniger94b}.  In this paper, we
summarize the recent result presented in \citet{revaz04} where warps
are spontaneously generated by bending instabilities \citep[see for
example][for a formal description of this instability]{fridman84} if a
substantial fraction of the dark matter resides in the disk of spiral
galaxies.

\section{Initial N-body models}

\begin{figure}
\begin{center}
\epsfig{file=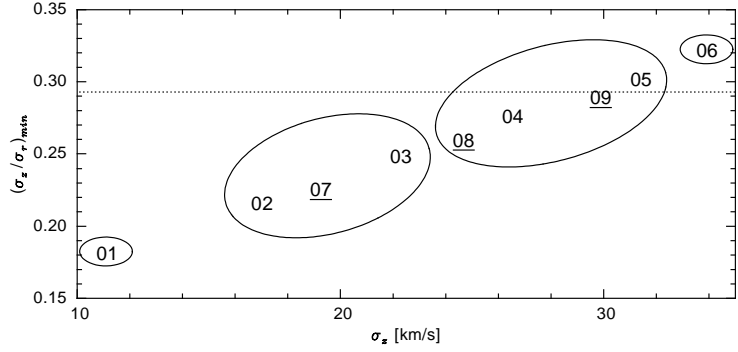,angle=-90,width=0.65\textwidth}
\caption{Ratio $\sigma_z/\sigma_R$ as a function of the vertical
  dispersion $\sigma_z$ at $R=15\,\rm{kpc}$.  The values are taken at
  the radius where $\sigma_z/\sigma_R$ is minimum.  The dotted
  line corresponds to Araki's limit.}
\label{fig01}
\end{center}
\end{figure}

In order to study the possible spontaneous vertical bending of spiral
galaxies, we have first studied the evolution of equilibrium N-body models
starting with different thicknesses, and without any dark halo.
Each model contains a bulge, an exponential disk and a dark matter
heavy disk parametrized by its thickness $h_{z0}$.  The initial
vertical velocity dispersion $\sigma_z$ is obtained by satisfying the
equilibrium solution of the Jeans equation, separately for each
component $i$.  The radial and azimuthal velocities dispersions
$\sigma_R$ and $\sigma_{\phi}$ are obtained from $\sigma_z$, using the
ratio between the epicycles frequencies $\kappa$ and $\nu$, and the
rotation frequency $\Omega$:
\begin{equation}
        \rho_i \, {\sigma_z}_i^2 = 
             \int_z^\infty\! dz \,\rho_i\, \partial_z \Phi,
	\qquad
        \frac{\sigma_{\phi}^2}{\sigma_R^2}= \frac{\kappa^2}{4\Omega^2}
	\qquad\textrm{and}\qquad
	\frac{\sigma_z^2}{\sigma_R^2}= \beta^2\,\frac{\kappa^2}{\nu^2}.
        \label{sigma}
\end{equation}
The free factor $\beta$ is chosen of order 1 to keep the
Savronov-Toomre radial stability parameter $Q$ around $1.5$.
Fig.~\ref{fig01} shows the Araki ratio $\sigma_z/\sigma_R$
\citep{araki85} as a function of the vertical dispersion $\sigma_z$.
Thin disks lie at lower left end while thick disks at the upper right
end.

\section{Simulations and results}

\begin{figure}
\epsfig{file=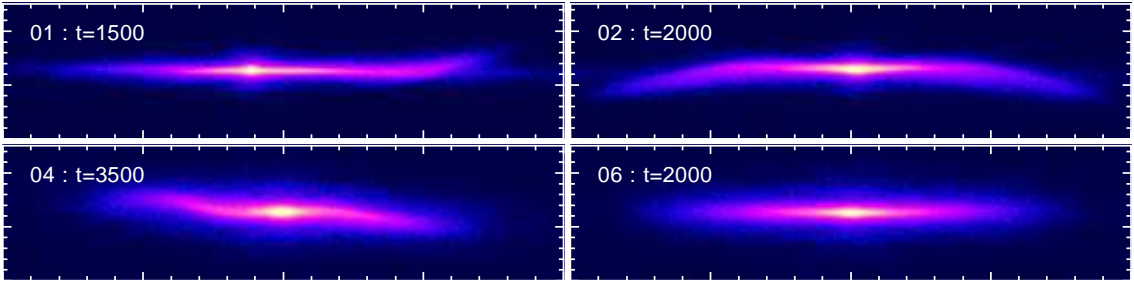,angle=-90,width=\textwidth}
\caption{Edge-on projections of the models 01, 02, 04, and 06.  Time
  in Myr is indicated at the upper left. The box dimensions are $100 \times
  25\,\rm{kpc}^2$}
\label{fig02}
\end{figure}

As expected from theory, the stability of the models depends strongly
on the precise thickness of the heavy disk.  In Fig.~\ref{fig01}, from left to
right, models can be divided into four groups. A characteristic snapshot
of each group is displayed in Fig.~\ref{fig02}.\\
1) Very thin models like model 01 ($h_{z0}<150\,\rm pc$) have a ratio
$\sigma_z/\sigma_R$ below $0.2$. They are highly unstable and generate
an asymmetric warp that extends up to  $z=4\,\rm{kpc}$ at $R=35\,\rm{kpc}$.\\
2) Models 02, 07 and 03 ($h_{z0}=150-250 \,\rm pc$) have still a ratio
$\sigma_z/\sigma_R$ well below the Araki limit. The bending instability occurs
during the first $2\,$Gyr. A spectacular axisymmetric bowl mode ($m=0$) grows
during about $1\,\rm{Gyr}$, before that $\sigma_z$ increases, which stabilizes
the disk.\\
3) In the models 08, 04, 09 and 05 ($h_{z0}=250-450 \,\rm pc$), the ratio
$\sigma_z/\sigma_R$ is almost critical with respect to Araki's limit. These
models  develop S-shaped warped modes ($m=1$). In the case of model 08, the
warp is long-lived and lasts more than $5.5\,\rm{Gyr}$.\\
4) The thickest model, 06 ($h_{z0}\ge 550\,\rm pc$) has a ratio
$\sigma_z/\sigma_R$ well above $0.3$. The disk remains stable.

\section{The influence of an extended dark halo}

\begin{figure}
\begin{center}
\epsfig{file=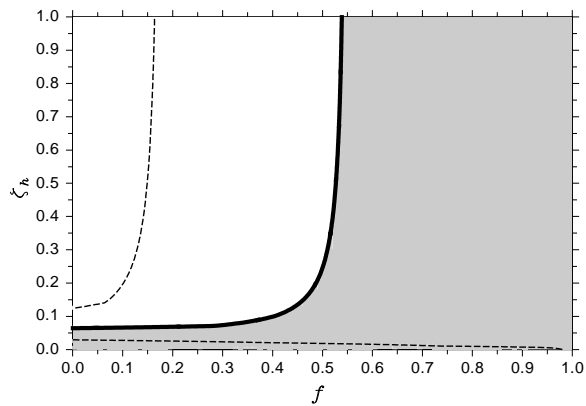,angle=-90,width=0.55\textwidth}
\caption{Stability of a realistic model $h_{z0}=250\,\rm{pc}$ against
  bending oscillations.  The radial minimum value of
  ${\sigma_R}/{\sigma_z}$ is plotted as a function of the dark
  matter fraction $f$ in the heavy disk and of the halo flattening
  $\zeta_h$. The solid line corresponds to the Araki critical value of
  $0.293$, the upper dashed line to the values of $0.4$, the bottom
  dashed line to $0.2$.  The shaded region corresponds to unstable
  models.}
\label{fig03}
\end{center}
\end{figure}

The previous models have been run without any outer dark halo, showing
that the initial $\sigma_z/\sigma_R$ ratio in the initial N-body runs
does determine the warp growth.  The influence of an external dark halo
on Araki's limit is then studied semi-analytically by embedding the
previous models in a Miyamoto-Nagai potential.  The heavy disk to
total dark mass ratio $f$ and the halo flattening $\zeta_h$
(iso-density axis ratio) characterize the semi-analytical models.  For
a realistic fixed disk thickness $h_{z0}=250\,\rm{pc}$,
Fig.~\ref{fig03} shows the ratio $\sigma_z/\sigma_R$ at its minimum
value along the radius, indicating the stability of the disk.
In the absence of the halo ($f=1$), the model is unstable. At the
opposite, when the heavy disk is replaced by a spherical halo
($\zeta_h=1$) of equal mass ($f=0$), the disk is stabilized, as long
as the halo flattening is realistic ($\zeta_h>0.5$). For very small
value of $\zeta_h$, the halo plays the role of the disk and the model
becomes unstable.  For a realistic $\zeta$, the halo flattening has
little effect on the stability.  In contrast the dark matter fraction
in the halo is well constrained: the disk remains stable as long as
the heavy disk to dark matter mass ratio is smaller than $0.5$, in
other words, bending instabilities appear only if the dark  
disk is heavier than the dark halo.

\section{Conclusions}

Our numerical N-body simulations have shown that heavy disks may be
subject to bending instabilities yielding S, U or asymmetric warps.
The shape and amplitude of these warps are similar to the observed
optical warps.  This scenario gives also an interesting constraint on
the halo mass in spiral galaxies. For realistic disk thickness,
bending instabilities may appear only if the halo is less massive than
about $0.3-0.4$ times the total galactic mass within the warped disk radius.

As discussed in more detail in \citet{pfenniger04}, the dissipational
behavior of the gas (not taken into account in the present
simulations) reduces the gas velocity perpendicular to the galactic
plane while, in the plane, higher values of velocities are maintained
by the dynamical heating due to instabilities like bars and spirals.
As a consequence, we expect that a galactic disk naturally maintains a
regime marginally unstable with respect to bending instabilities. 
Like spiral arms, warps can be repeatedly excited as long as the 
dissipative component, the gas, exists.

\end{document}